\documentclass[twocolumn,prc,aps,nofootinbib,showkeys,showpacs]{revtex4}
\usepackage[latin9]{inputenc}
\usepackage{color}
\usepackage{amsmath}
\usepackage{amsthm}
\usepackage{amssymb}
\usepackage{graphicx}

\makeatletter

\usepackage{epsfig}
\usepackage{amsthm}
\usepackage{tikz}
\usetikzlibrary{quantikz}

\usepackage{bm}

\usepackage{color}

\makeatother

\begin{document}

\title{ Calculation of generating function in many-body systems with quantum computers: technical challenges and use in hybrid quantum-classical methods\footnote{Note that the article had initially the title "Predicting ground state, excited states and long-time evolution of many-body systems from short-time evolution on a quantum computer", we decided to change the title in view of the difficulty to achieve good precision in the estimation of $\langle H^K \rangle$ using the finite difference method. In general, good precisions can be achieved for low $K$ ($K<7$ in the present model case) values but degrades as $K$ increases. Alternative approaches that avoid this difficulty is explored in Ref. \cite{Rui21}.}}
\author{Edgar Andres Ruiz Guzman }
\email{ruiz-guzman@ijclab.in2p3.fr}

\affiliation{Universit\'e Paris-Saclay, CNRS/IN2P3, IJCLab, 91405 Orsay, France}
\author{Denis Lacroix }
\email{denis.lacroix@ijclab.in2p3.fr}

\affiliation{Universit\'e Paris-Saclay, CNRS/IN2P3, IJCLab, 91405 Orsay, France}
\date{\today}
 
\begin{abstract}
The generating function of a Hamiltonian $H$ is defined as $F(t)=\langle e^{-itH}\rangle$,
where $t$ is the time and where the expectation value is taken on a given initial quantum state.
This function gives access to the different moments
of the Hamiltonian $\langle H^{K}\rangle$ at various orders $K$. The real and imaginary parts of $F(t)$ can be respectively evaluated 
on quantum computers using one extra ancillary qubit with a set of measurement for each value of the time $t$. 
The low cost in terms of qubits renders it very attractive in the near term period where 
 the number of qubits is limited. Assuming that the generating function can be precisely 
 computed using quantum devices, we show how the information content of this function can be used 
 a posteriori on classical computers to solve quantum many-body problems.  Several methods of classical 
 post-processing are illustrated with the aim to predict approximate
ground or excited state energies and/or approximate long-time
evolutions. This post-processing can be achieved using methods
based on the Krylov space and/or on the $t$-expansion approach that
is closely related to the imaginary time evolution. Hybrid quantum-classical calculations
are illustrated in many-body interacting systems using the pairing
and Fermi-Hubbard models. 
\end{abstract}


\maketitle

\section{Introduction}

With recent advances in the development of quantum computing (QC)
platforms, the possibility of exploiting quantum devices for realistic
simulations of complex quantum systems, as suggested by Feynman \cite{Fey82},
is becoming a reality (see for instance \cite{Aru19,Ale21}). Nowadays,
quantum simulations are possible, but the quantum noise and decoherence
significantly limit the number of operations that could be performed
efficiently on existing platforms. This is what nowadays is called
the NISQ (Noisy Intermediate-Scale Quantum) era \cite{Pre18} where
simulations on quantum computers are possible but the algorithms and
tasks should adapt to noise. Because of this noise, many standard
algorithms cannot be used in actual QC devices while others appear
particularly suited in the NISQ context. In the present work, we are
interested in simulating complex quantum systems. In this context,
a typical example of NISQ "friendly" strategy is the use of Variational
optimizers using Hybrid quantum-classical architectures where the
optimization is made with a classical computer \cite{End21,Cer21}.
These developments have given a strong impulse to the use of quantum
computers for calculating complex quantum many-body systems in different
fields of physics \cite{Lan10,Bab15,OMa16,Col18,Hem18,Mac18,Dum18,Lu19,Rog19,Du20,Klc18,Klc19,Ale19,Lam19}.
For recent reviews on the subject see for instance \cite{Mcc17,Fan19,Cao19,McA20,Bau20,Bha21}.

Here, we explore a different hybrid strategy to simulate quantum systems.
Our starting hypothesis is that the QC can simulate the evolution
of a quantum system, at least approximately, over a restricted time
interval $[0,t_{{\rm max}}]$. 
The method we propose can be seen as a natural
generalization of the one used in Refs. \cite{Kni07,Rog20} where
the expectation value of an Hermitian operator $O$ is replaced by
the evolution of the operator $e^{-itO}$ over a short time interval.
Here, we use the standard concept of generating function. The generating function is already used
implicitly in the context of quantum computing in the quantum phase
estimation (QPE) algorithm \cite{Nie02}. It was exploited recently
in Ref. \cite{Lac20} to restore symmetries in many-body systems.
However, in this case, the circuit is too deep to be simulated in
the NISQ period. Here, we show that the generating
function (GF) can be obtained using a single ancillary qubit. 
Precise estimates of the GF gives a priori access to 
set of moments $\langle H^{K}\rangle$ with $K\le M$ using
the GF. This technique was already discussed in Ref. \cite{Pen21,Cla21} in combination with variational principles 
and noted as a possible tool for the NISQ period. 
We give illustration here of methods where the moments can be used in a second step for a post-processing
on a classical computer to study the static and dynamical properties
of complex systems.

\section{Generating function on quantum computers and $\langle H^{K}\rangle$
estimates}

\label{sec:generating}

The generating function is a standard concept of classical probability
and statistical theory. We recall briefly here how this concept can be
exploited in quantum systems \cite{Bal07}. We
consider a system described by a density $\rho$. In the following,
we assume implicitly that ${\rm Tr}(\rho)=1$. If
the system is in a pure state, the density can be written as $\rho=|\Phi \rangle\langle\Phi |$.

For a given operator $O$, we can define the generating function as:
\begin{eqnarray}
F(\gamma) & = & {\rm Tr}\left(e^{\gamma O}\rho \right),\label{eq:gen}
\end{eqnarray}
where $\gamma$ is a complex number. The interest of the generating
function is that its knowledge gives access to the different non-centered
moments $\langle O^{K}\rangle={\rm Tr}(O^{K}\rho)$ associated
to the density $\rho$. Indeed, expanding the exponential, we
deduce: 
\begin{eqnarray}
F(\gamma) & = & 1+\gamma\langle O\rangle+\frac{\gamma^{2}}{2!}\langle O^{2}\rangle+\cdots\label{eq:taylor0}
\end{eqnarray}
that corresponds to the Taylor expansion of $F(\gamma)$ with the
condition: 
\begin{eqnarray}
\left.\frac{d^{K}F(\gamma)}{d\gamma^{K}}\right|_{\gamma=0} & = & \langle O^{K}\rangle.\label{eq:taylorder}
\end{eqnarray}
Until now, we have not specified $\gamma$. Our aim is to estimate
the generating function on a quantum computer. Since quantum computers
are convenient to perform unitary evolutions, it is suitable to take
$\gamma=-it$. Then, if $O$ is Hermitian, $e^{-itO}$ is a unitary
operator.

We will focus our attention here on the case where $O$ identifies
with a Hamiltonian denoted by $H$. Assuming $\hbar=1$, the operator
entering in Eq. (\ref{eq:gen}) is simply the propagator
in time $U(t)=e^{-itH}$. In practice, the simulation of non-unitary
(but Hermitian) operators, such as the Hamiltonian or its powers, on
a quantum computer is a much more complicated task than performing
$U(t)$ itself (see for instance the discussion in \cite{Ber15}).
The GF provides a practical tool to estimate the
expectation values of such non-unitary operators while performing
only unitary operations.

The GF is already used explicitly or
implicitly in the quantum computing context. For instance, the quantum
phase-estimation (QPE) approach \cite{Nie02,Hid19,Fan19,Ovr03,Ovr07}
applied to an operator $U_{S}=e^{2\pi iS}$ is actually computing
the generating function associated to the operator $S$ on a set of
ancillary qubits prior to performing the quantum inverse Fourier transform
to obtain the probability distribution of the eigenstates of $S$.
The GF is also a key ingredient of the time-series
method discussed in Ref. \cite{Som19}.

Our strategy in the present work is to assign to the quantum computer
solely the task of computing the GF, even on a restricted
interval of time, with the additional constraint to minimize
the number of ancillary qubits. The generating function is then transmitted
as input to a classical computer for post-processing. We will give
below several illustrations of such post-processing.

\begin{figure}
\includegraphics[width=1.0\linewidth]{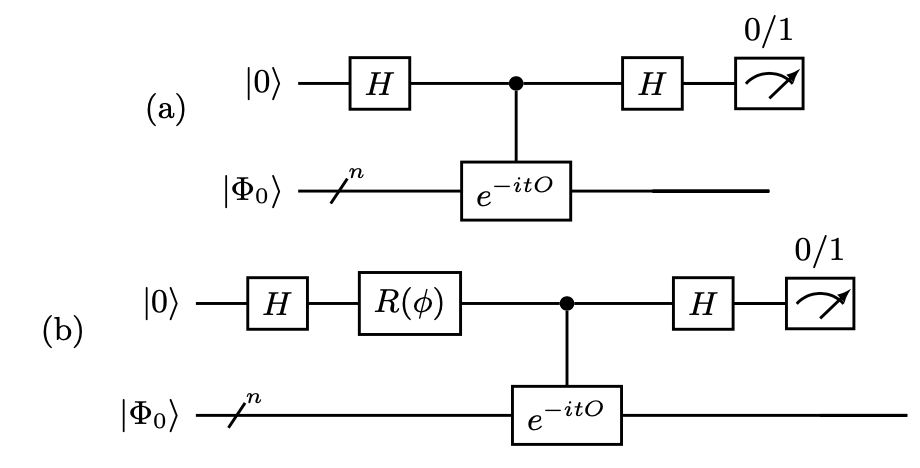}
 \caption{Illustration of the (a) Hadamard test and (b) modified Hadamard test
that are used in the present work to compute respectively the real
and imaginary parts of the GF for a given
Hermitian operator $O$. In this circuit, $H$ is the standard Hadamard
gate while $R(\phi)$ corresponds to the phase gate where the angle
is set to $\phi=-\pi/2$. In the case (a), the probability to measure
0 or 1 on the ancillary qubit verifies $p_{0}-p_{1}={\rm Re}\{F(t)\}$
while in the case (b) we have $p_{0}-p_{1}={\rm Im}\{F(t)\}$. The
circuits shown here and in the following figures have been made using
the quantikz package of Ref. \cite{Kay18}. }
\label{fig:hadgen} 
\end{figure}

On a quantum computer, the GF can be obtained by
adding a single register qubit to the ones used for the system itself. For a
given value of $t$, the real and imaginary parts of $F(t)$ are 
obtained using the standard Hadamard test or the modified Hadamard
test, as shown respectively in panels (a) and (b) of Fig. \ref{fig:hadgen}, by measuring the additional qubit.
Note that, a set of measurements is required for each values of the
time. Illustrations of generating functions are given below for interacting
fermions.

\subsection{Illustration of the method}

To illustrate the method, we consider two different Hamiltonians that
are standardly used to test many-body approaches, namely the pairing
Hamiltonian and the one dimensional Fermi-Hubbard model. In both
cases, we have used the Jordan-Wigner transformation (JWT) \cite{Jor28,Lie61,Som02,See12,Dum18,Fan19}
to map the Hamiltonian written in second quantization into a set of
interacting qubits. We take the following specific convention for
the mapping. Assuming a set of fermion creation/annihilation operators
$(a_{j}^{\dagger},a_{j})$, we map these operators into qubits gates
such that 
\begin{eqnarray}
\left\{ \begin{array}{l}
a_{j}^{\dagger}\longrightarrow A_{j}^{+}=Z_{j-1}^{<} \otimes Q_{j}^{+}\\
\\
a_{j}\longrightarrow A_{j}=Z_{j-1}^{<}\otimes  Q_{j} 
\end{array}\right.,\label{eq:jwt}
\end{eqnarray}
with the definitions
\begin{eqnarray}
Q_{j}=\frac{1}{2}\left(X_{j}+iY_{j}\right),~~~Q_{j}^{+}=\frac{1}{2}\left(X_{j}-iY_{j}\right).
\end{eqnarray}
Here $(X_{j},Y_{j},Z_{j})$ are the standard Pauli matrices acting
on the qubit $j$. We add to these operators the identity operator
$I_{j}$. In the equation (\ref{eq:jwt}), we have defined the quantity
$Z_{j-1}^{<}$ as 
\begin{eqnarray*}
Z_{j-1}^{<} & = & \bigotimes_{k=0}^{j-1}(-Z_{k}).
\end{eqnarray*}
With this convention we have for instance $Q_{j}^{+}|0_{j}\rangle=|1_{j}\rangle$
and $a_{k}^{\dagger}a_{k}\rightarrow N_{k}=(I_{k}-Z_{k})/2$. Basic
aspects related to the quantum simulation of both model Hamiltonians
considered here are summarized below.

\subsubsection{Fermi-Hubbard model}

The Fermi-Hubbard model is a widely used schematic model to describe
interacting fermions on a lattice \cite{Jak98,Gre02}. The Hubbard
Hamiltonian  was already simulated on quantum computers in Ref.
\cite{Wec15,Jia18}. We consider here the one-dimension Fermi-Hubbard
model with sharp boundary conditions. The Hamiltonian describes a
set of $N$ fermions with spins on a set of $M$ lattice sites which
are labeled as $i=0,1,\dots,M-1$. This Hamiltonian is written as
$H=H_{J}+H_{U}$, where $H_{J}$ and $H_{U}$ are the hopping and
interaction terms respectively given by: 
\begin{eqnarray*}
H_{J} & = & -J\sum_{i,\sigma}(a_{i+1,\sigma}^{\dagger}a_{i,\sigma}+a_{i,\sigma}^{\dagger}a_{i+1,\sigma}),\\
H_{U} & = & +U\sum_{i}\hat{n}_{i,\uparrow}\hat{n}_{i,\downarrow},
\end{eqnarray*}
with $n_{i,\sigma}=a_{i,\sigma}^{\dagger}a_{i,\sigma}$ and $\sigma=\{\uparrow,\downarrow\}$.
In order to apply the JWT mapping, it is convenient to organize the
qubits as follows. Spin-up single-particle states indexed as
$i=0,\dots,M-1$ are associated with qubits labeled with $\alpha=0,\dots,M-1$.
Particles with spin-down indexed as $i=0,\dots,M-1$ are associated
to qubits $\alpha=M,\dots,2M-1$.
With this, we obtain the mapping (with proper account for the boundary
conditions): 
\begin{eqnarray*}
H_{J} & = & J\sum_{\alpha=0,\alpha\neq M-1}^{2M-2}\left[Q_{\alpha+1}^{+}Q_{\alpha}+{\rm h.c.}\right],
\end{eqnarray*}
together with 
\begin{eqnarray}
H_{U}=\frac{U}{4}\sum_{\alpha=0,M-1}\left[I_{\alpha}-Z_{\alpha}\right]\left[I_{\alpha+M}-Z_{\alpha+M}\right].
\end{eqnarray}

The generating function evaluation with the circuits presented in
Fig. \ref{fig:hadgen} requires to perform the time-evolution operator.
For its implementation, we simply use the Trotter-Suzuki method \cite{Tro59,McA20}.
The time interval $[0,t]$ is divided into small intervals $\Delta t$.
For small enough time interval, we have: 
\begin{eqnarray*}
U(\Delta t)=e^{-i\Delta tH}\simeq e^{-i\Delta tH_{J}}e^{-i\Delta tH_{U}}\equiv U_{J}(\Delta t)U_{U}(\Delta t).
\end{eqnarray*}
The propagators $U_{J}$ can be further decomposed as: 
\begin{eqnarray}
U_{J}(\Delta t) & = & \prod_{\alpha}e^{-iJ\Delta t\left[Q_{\alpha+1}^{+}Q_{\alpha}+{\rm h.c.}\right]},\nonumber \\
 & = & \prod_{\alpha}\begin{pmatrix}1 & 0 & 0 & 0\\
0 & \cos\left(\lambda\right) & -i\sin\left(\lambda\right) & 0\\
0 & -i\sin\left(\lambda\right) & \cos\left(\lambda\right) & 0\\
0 & 0 & 0 & 1
\end{pmatrix}_{\alpha,\alpha+1}
\end{eqnarray}
with $\lambda=\Delta tJ$. To obtain the matrix form, standard manipulation
of Pauli matrices is used. Note that the index on the matrix indicates
that the matrix acts on the two qubits $\alpha$ and $\alpha+1$.

\begin{figure}
\includegraphics[width=0.7\linewidth]{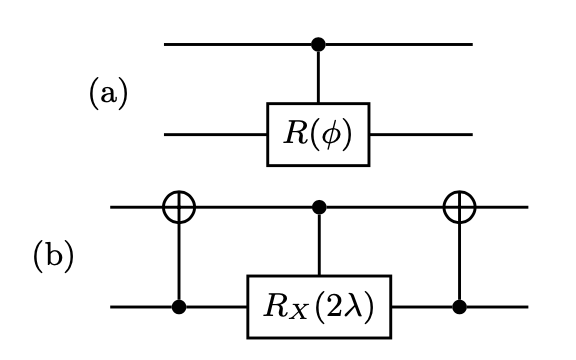}
 \caption{Circuits used to simulate the Hubbard model. The circuit (a) simulates
the interaction term $H_{U}$ where $R(\phi)$ is the unitary phase
operator with $\phi=-\Delta tU$. Circuit (b) simulates a short time-step
evolution of the hopping term $H_{J}$ where $R_{X}(2\lambda)=e^{-i\lambda X}$
and where $\lambda=J\Delta t$.}
\label{fig:circuj} 
\end{figure}

For the interaction propagator we have 
\begin{eqnarray}
U_{U}(\Delta t) & = & \prod_{\alpha}e^{-iU\Delta t\left[I_{\alpha}-Z_{\alpha}\right]\left[I_{\alpha+M}-Z_{\alpha+M}\right]},\nonumber \\
 & = & \prod_{\alpha}\begin{pmatrix}1 & 0 & 0 & 0\\
0 & 1 & 0 & 0\\
0 & 0 & 1 & 0\\
0 & 0 & 0 & e^{-i\Delta tU}
\end{pmatrix}_{\alpha,\alpha+M}
.
\end{eqnarray}
We recognize in the last expression the controlled phase-shift gate
with phase $\phi=-\Delta tU$. The two circuits that simulate $U_{U}$
and $U_{J}$ are displayed in panels (a) and (b) of Fig. \ref{fig:circuj}.

\subsubsection{Pairing Hamiltonian}

As a second illustration, we will also consider the pairing Hamiltonian
\cite{Von01,Zel03,Duk04,Bri05} that is standardly used in the context
of nuclear physics or small superconducting systems. This Hamiltonian
has already been used on QC in Refs. \cite{Ovr03,Ovr07} and more
recently in Refs. \cite{Lac20,Kha21}. We write this Hamiltonian
as: 
\begin{eqnarray}
H & = &  \sum_p \varepsilon_p N_p + g \sum_{pq} P^\dagger_p P_q  \equiv H_{\varepsilon}+H_{g}. 
\end{eqnarray}
Introducing the notation $(a_{p}^{\dagger},a_{\bar{p}}^{\dagger})$
as the creation operators of time-reversed single-particle states. The different operators
are defined as: 
\begin{eqnarray*}
\hat{N}_{p} & = & a_{p}^{\dagger}a_{p}+a_{\bar{p}}^{\dagger}a_{\bar{p}},\\
\hat{P}_{p}^{\dagger} & = & a_{p}^{\dagger}a_{\bar{p}}^{\dagger}.
\end{eqnarray*}
These operators correspond respectively to the pair occupation, and
to the pair creation operators. In this model, time-reversed single-particle
states are degenerated with energies $\widetilde{\varepsilon}_{p}=\varepsilon_{p}+g_{pp}/2$,
where the $g_{pp}/2$ term is added to compensate from the shift induced
by scattering of each pair by itself in the $H_{g}$ term (case $p=q$).

The mapping from fermions to qubit of the pairing problem can be made
in different ways. In the most general situation, one can follow the
standard JWT where one particle corresponds to one qubit. This was
done for instance in Ref. \cite{Ovr07} or in \cite{Lac20}. The method to map 
fermions to qubits used in these works is general and can treat the case of system with odd or even particle numbers.   
We are interested here only in systems with even number of
particles with the particularity that there is no broken pairs (seniority
zero scheme \cite{Bri05}), one can then directly map each pair operator
$P_{p}^{\dagger}$ into a single qubit. This was done in Ref. \cite{Kha21}
with the advantage to reduce the number of qubits needed to describe
the system. Here, we use the latter strategy. Following Ref. \cite{Kha21},
the Hamiltonian in the qubits space is written as: 
\begin{eqnarray}
H & = & \sum_{p} \varepsilon_{p}\left[1-Z_{p}\right]-\frac{1}{2}\sum_{p>q}g_{pq}\left[X_{p}X_{q}+Y_{p}Y_{q}\right].
\end{eqnarray}
We apply the Trotter-Suzuki method to this Hamiltonian and denote
by $U_{\varepsilon}(\Delta t)$ and $U_{g}(\Delta t)$ the propagator
associated respectively to $H_{\varepsilon}$ and $H_{g}$ for small
time-step evolution $\Delta t$. For the one-body part of the Hamiltonian,
we have: 
\begin{eqnarray}
U_{\varepsilon}(\Delta t) & = & \prod_{p}\exp\left(-i\Delta t\varepsilon_{p}\left[1-Z_{p}\right]\right),\nonumber \\
 & = & \prod_{p}\left(\begin{array}{cc}
1 & 0\\
0 & \exp\left(-2i\varepsilon_{p}\Delta t\right)
\end{array}\right)\equiv\prod_{p}R(\phi_{p}),
\end{eqnarray}
where $R(\phi_{p})$ is the unitary phase-gate operator with $\phi_{p}=-2\varepsilon_{p}\Delta t$.
For the interaction part, we have: 
\begin{eqnarray}
U_{g}(\Delta t) & = & \prod_{p>q}\exp\left(+ig_{pq}\frac{\Delta t}{2}\left[X_{p}X_{q}+Y_{p}Y_{q}\right]\right)\nonumber \\
 & = & \prod_{p>q}\begin{pmatrix}1 & 0 & 0 & 0\\
0 & \cos(\lambda_{pq}) & i\sin(\lambda_{pq}) & 0\\
0 & i\sin(\lambda_{pq}) & \cos(\lambda_{pq}) & 0\\
0 & 0 & 0 & 1
\end{pmatrix}. 
\end{eqnarray}
where we have defined $\lambda_{pq}=g_{pq}\Delta t$. We recognize
the same matrix form as for the $H_{J}$ term that could be simulated
using the circuit shown in panel (b) of Fig. \ref{fig:circuj}.

\begin{figure}[htbp]

\includegraphics[width=1.0\linewidth]{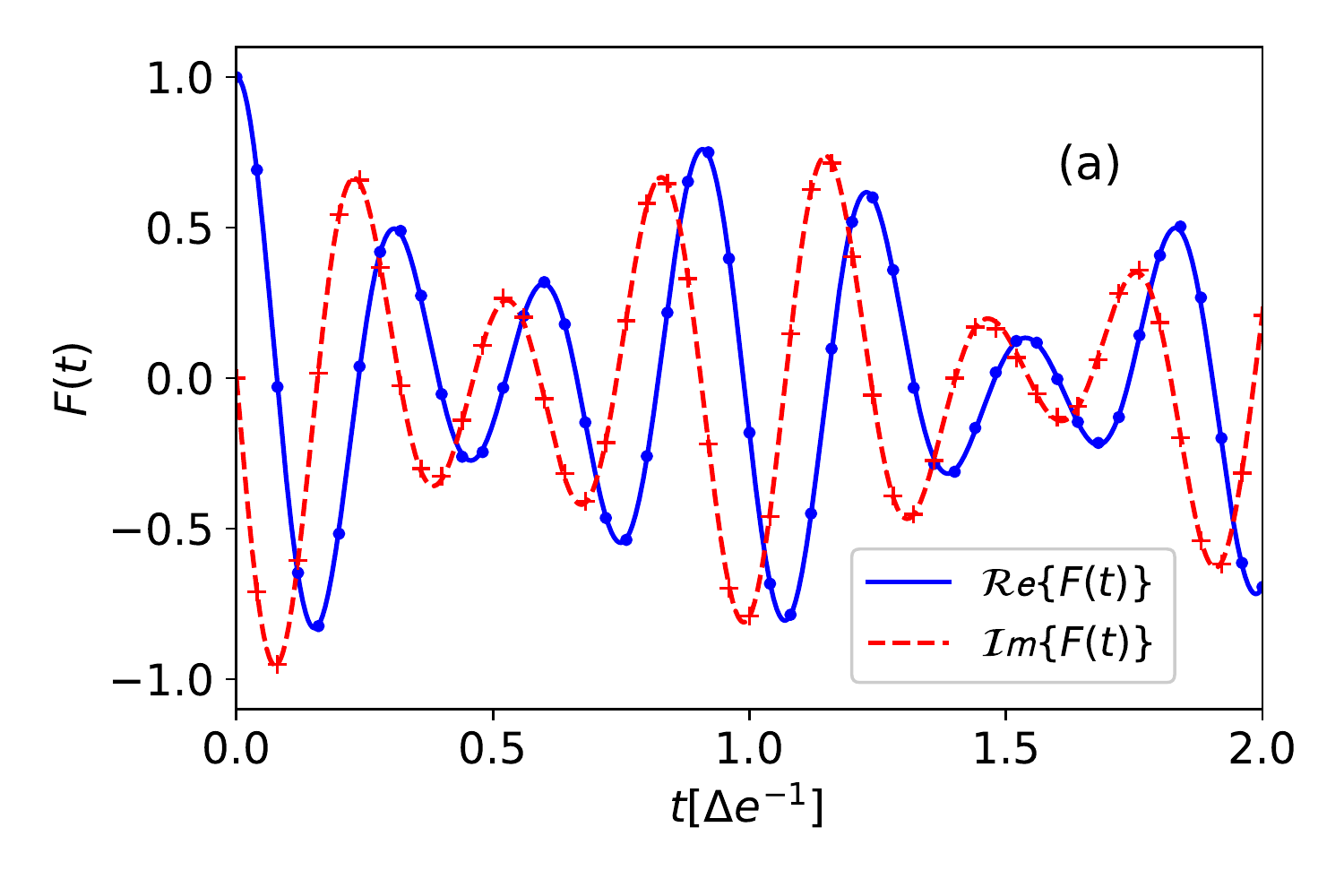}
\includegraphics[width=1.0\linewidth]{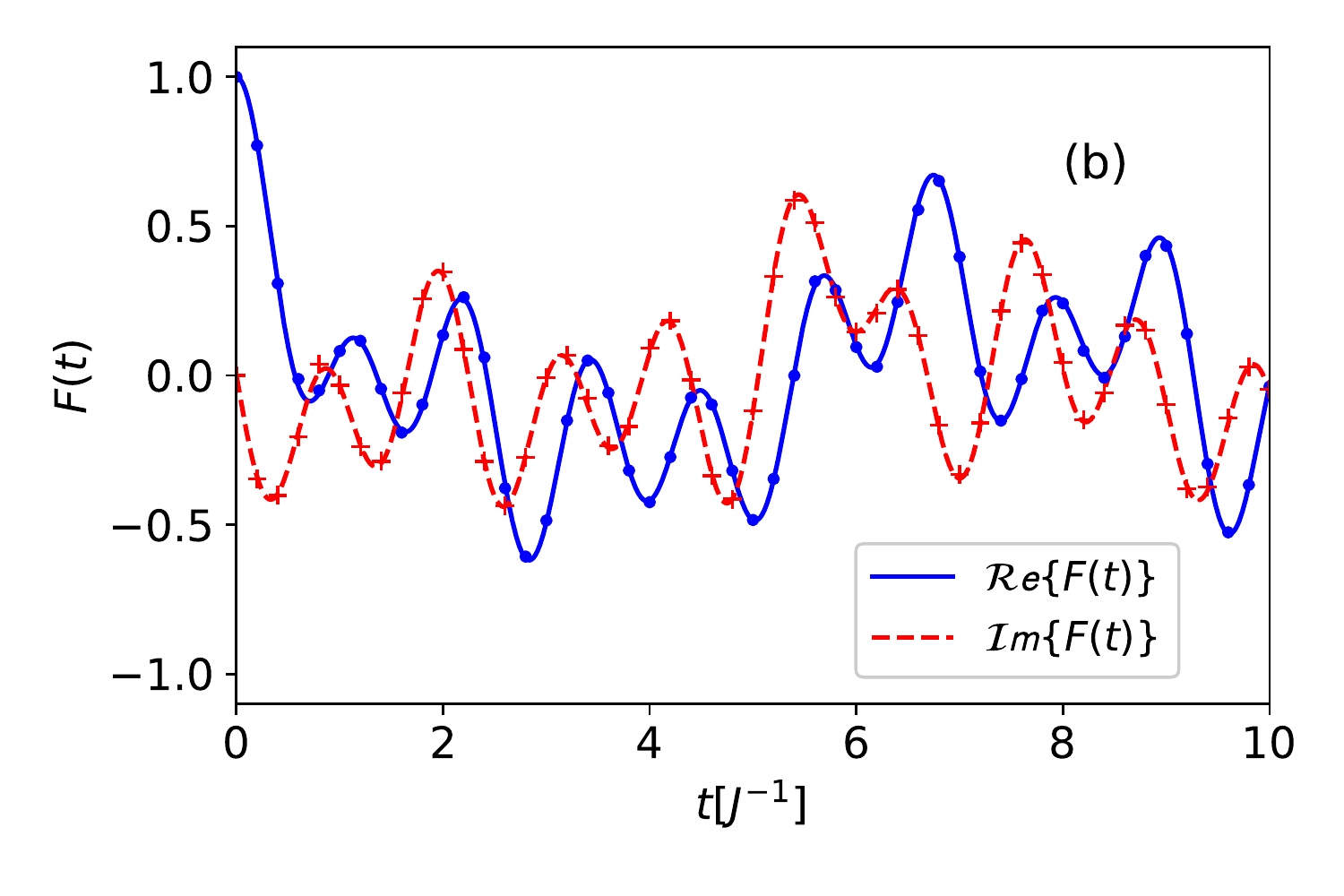} 

\caption{Panel (a): Illustration of the real (blue solid line) and imaginary
(red dashed line) parts of the generating function obtained on a classical
computer for the pairing model with $N=4$ pairs on $M=8$ doubly
degenerated single-particle levels with energy $\varepsilon_{p}=p\Delta e$
(with $p=1,\dots,M$) and $g_{pq}={\rm cte}=g/\Delta e=1$. The initial
condition corresponds to the Slater determinants where the 4 lowest
single-particle energies are occupied. In panel (b), we show the same
quantities obtained for the Fermi-Hubbard model with $N=4$ particles
on $M=4$ sites for the parameters $U/J=1$. The initial condition
corresponds to a spin-saturated case where the initial state is an
average over the $C_{4}^{2}=6$ Slater determinants where pairs of
particles with opposite spins occupy randomly 2 sites among the 4
possibilities. Results displayed with symbols in both panels correspond
to calculation obtained with the Qiskit-qasm QC software \cite{Abr19}
and the circuits of Fig. \ref{fig:hadgen}. Each point in these
figures is obtained by averaging over $10^{4}$ measurements. }
\label{fig:genhubpair} 
\end{figure}

\subsubsection{Illustration of generating function and Hamiltonian moments obtained
by quantum computation }

With the use of the Hadamard test and its modified version, together with
the different circuits required to perform the time-evolution, we
have now all ingredients to extract the real and imaginary part of the generating function $F(t)$
with only one extra ancillary qubit.

We show in Figure \ref{fig:genhubpair} the real and imaginary parts
of the generating function obtained in the two model cases. The lines
correspond to the GF obtained on a classical computer directly by diagonalization of the Hamiltonian. The symbols
are the results obtained with the QC simulator using the two circuits
shown in Fig. \ref{fig:hadgen}. Each points reported in this figure are calculated 
by averaging $10^{4}$ events using the perfect quantum
computer (IBM Qiskit toolkit with qasm \cite{Abr19}). Not surprisingly,
since the emulator simulates a perfect QC without noise, the results
obtained on the quantum and classical computers perfectly coincide with each other. The
only condition is to perform sufficient measurements and to use a
numerical time-step $\Delta t$ small enough to insure that the Trotter-Suzuki
approximation is valid. We used here $\Delta t.J=0.02$ and $\Delta t.\Delta e=0.002$
for the Fermi-Hubbard and pairing model respectively.

\subsection{Physical content of the generating function}

\label{sec:ftstrength}

The knowledge of the response function at all time gives access to
the spectral properties of the Hamiltonian. Indeed, if we introduce
a complete set of eigenstates $|\alpha\rangle$ of the Hamiltonian
with energy $E_{\alpha}$, we have: 
\begin{eqnarray}
F(t) = \sum_{\alpha}e^{-itE_{\alpha}}\langle\alpha|\rho_{0}|\alpha\rangle,=\sum_{\alpha}e^{-itE_{\alpha}}|\langle\alpha|\Phi_{0}\rangle|^{2},
\label{eq:edecomp}
\end{eqnarray}
where the last identity holds for a pure initial state $|\Phi_{0}\rangle$.
Knowing the generating function for all times gives both 
the eigenstates energies $E_{\alpha}$ and the amplitudes $|\langle\alpha|\Phi_{0}\rangle|^{2}$.
The Fourier transform of the GF, denoted by $\widetilde{F}(\omega)$,
also related to the strength or response functions, verifies: 
\begin{eqnarray}
\widetilde{F}(\omega) & \propto & \sum_{\alpha}\delta(\omega-E_{\alpha})|\langle\Phi_{0}|\alpha\rangle|^{2}.\label{eq:strength}
\end{eqnarray}
Such response function can be computed directly within the quantum
phase-estimation technique \cite{Nie02} using a set of ancillary
qubits or using only one ancillary qubit as proposed in the present
work or in Ref. \cite{Som19}.


A second interesting property of the generating
function is its connection with the moments, see Eq. (\ref{eq:taylorder}).   
For the specific case $\gamma=-it$
and $O=H$, we have the relationship: 
\begin{eqnarray}
\langle H^{K}\rangle & = & i^{K}\left.\frac{d^{K}F(t)}{dt^{K}}\right|_{t=0}.\label{eq:deriv}
\end{eqnarray}
So that a perfect knowledge of the generating function for all times $t$, gives a priori access 
to the expectation value of the moments $H^K$ calculated for the initial state.   

\subsection{Critical discussion of the extraction of the moments from the generating function}

In the original version of the present article, we proposed to use the finite difference method for the estimate 
of the left-hand side of Eq. (\ref{eq:deriv}). In practice, one could indeed approximately access the different values
of $\langle H^{K}\rangle$ by replacing the derivatives by their
finite difference expressions (a comprehensive list of finite difference
coefficients with various level of accuracy to estimate the derivatives
are given for instance in \cite{For88,Fdi20}). The finite difference method (FDM) is indeed 
adequate to obtain rather precise values for the first few moments. This is, for instance, 
the practical method used in Ref. \cite{Ger17,Mit18,Sch19} to simulate the first derivative of the objective function
in the context of quantum machine learning. \\
The precision on the moments however degrades when the order $K$ increases. Noteworthy, the methods discussed 
below (Pad\'e or Krylov based) 
requires rather precise determination of the moments. We have made significant efforts to optimize both the time step 
and the number of points used in the finite-difference. Our conclusion is that the FDM, even assuming noiseless quantum computers, 
cannot reach sufficient accuracy to compute $\langle H^K \rangle$ as $K$ increases. In the model Hamiltonian considered here, that are relatively simple compare to more realistic Hamiltonians in quantum chemistry or nuclear physics, relatively good accuracy 
can be achieved with the FDM for moments up to $K=10$ to $15$ depending on the interaction strength. Even if the order of this moments are already quite high,  we have observed that a small error on the estimated moments can impact significantly the precision on the post-processing. 


Besides the FDM approach, we made extensive tests of
polynomial interpolation (standard and Chebyshev) to obtain high precision on the moments. Again, polynomial methods 
are able to achieve reasonable precision of first moments but are not accurate enough for high $K$ values.           

The only approaches that were able to achieve global convergence of all moments with sufficient accuracy are those based 
on the Fourier transform of the generating function. Indeed, performing the Fourier gives access to approximation of the 
components of the initial 
state on the eigenstates, denoted by $p_\alpha$, 
as well as to a set of approximate eigenenergies $\widetilde E_\alpha$ (see Eq. (\ref{eq:strength}))\footnote{We do not recall here basic ingredients of the Fourier Transform technique for time-dependent signal processing that could be found in many textbooks. }.  
Then, from this information, one can simply 
obtain approximation of $\langle H^K \rangle$ using the formula $\sum_\alpha \widetilde E_\alpha^K p_\alpha$. In the absence of noise on the signal, very good approximation of 
the moments to any order can be obtained provided that the time-step used is sufficient small to resolve the largest eigen-energy and the time
interval $t_{\rm max}$ is sufficiently large to achieve a good energy resolution. The necessity to use Fourier transform requires to compute
the generating function for many time steps over rather long time. As a consequence, this significantly increases the effort required to compute the GF on the quantum computer. This aspect, that we seriously underestimated in the first version of this work, render the approach less attractive, especially compared to the quantum-phase-estimation approach that is also based on the Fourier technique. 

Despite this difficulty, we give below some illustrations of some possible post-processing assuming that 
the moments can be computed with good precision.

\section{Illustration of applications}

In this section, we assume that we have obtained a set of moments
of the Hamiltonian up to a given, yet limited, order $L$ and illustrate
how this information can be used in a second step for post-processing
on a classical computer.

\subsection{$t$-expansion approach for the ground state energy}

As a first illustration, we consider the $t$-expansion technique
introduced in Ref. \cite{Hor84} and considered more recently in \cite{Sek20}
in the context of quantum computing. One of the goal of the approach
is to obtain the ground state energy denoted by $E_{{\rm GS}}$. In
the following, we denoted by $|\Psi_{{\rm GS}}\rangle$ the ground
state wave-function.

Given an initial state $|\Phi_{0}\rangle$, our objective is to perform
the imaginary-time evolution of this state up to a given time $\tau$,
leading to the state 
\begin{eqnarray*}
|\Psi(\tau)\rangle & \equiv & \frac{e^{-\tau/2H}}{\sqrt{\langle e^{-\tau H}\rangle}}|\Phi_{0}\rangle.
\end{eqnarray*}
We know that, whatever the initial state $|\Phi_{0}\rangle$, if initially
$\langle\Phi_{0}|\Psi_{{\rm GS}}\rangle\neq0$, then $|\Psi(\tau)\rangle$
will converge to the ground state $|\Psi_{{\rm GS}}\rangle$.
We then have: 
\begin{eqnarray}
E_{{\rm GS}} & = & \lim_{\tau\rightarrow\infty}\langle\Psi(\tau)|H|\Psi(\tau)\rangle=\lim_{\tau\rightarrow\infty}E(\tau).
\end{eqnarray}
The key aspect underlined in Ref. \cite{Hor84} was to show that
the convergence of the energy towards the ground state is directly
connected to the moments of the Hamiltonian estimated at initial time.

This could be shown by noting that: 
\begin{eqnarray}
E(\tau) & = & \frac{\langle He^{-\tau H}\rangle}{\langle e^{-\tau H}\rangle}=-\frac{d}{d\tau}\ln\langle e^{-\tau H}\rangle
,
\end{eqnarray}
where the expectation values are taken on the initial state $|\Psi_{0}\rangle$.
We recognize in the last expression the generating function $Z(\tau)$
of the cumulants of $H$. More precisely, we have the relationship:
\begin{eqnarray}
Z(\tau) & = & \ln\langle e^{-\tau H}\rangle=\sum_{K=0}^{+\infty}\frac{(-\tau)^{K}}{K!}\kappa_{K},\label{eq:taylorz}
\end{eqnarray}
where $\kappa_{K}$ is the cumulant of order $K$ of the Hamiltonian
that are calculated from the moments of orders lower or equal to $K$
with the initial state. For the sake of completeness, we recall the 
useful recurrence relation:
\begin{eqnarray}
\kappa_{n} &=& \langle H^{n}\rangle - \sum_{k=1}^{n-1}\begin{pmatrix}n-1\\
k-1
\end{pmatrix}\kappa_{n}\langle H^{n-k}\rangle ,\nonumber
\end{eqnarray}  
that could be used iteratively with the condition $\kappa_1 =\langle H\rangle$.

Having the set of moments up to a given order informs us on the
value of $E(\tau)$ over a certain imaginary time interval $[0,\tau_{max}]$.
This interval depends only on the initial state that determines the
moment values as well as on the number of available moments.

\begin{figure}[htbp]
\begin{centering}
\includegraphics[width=1\linewidth]{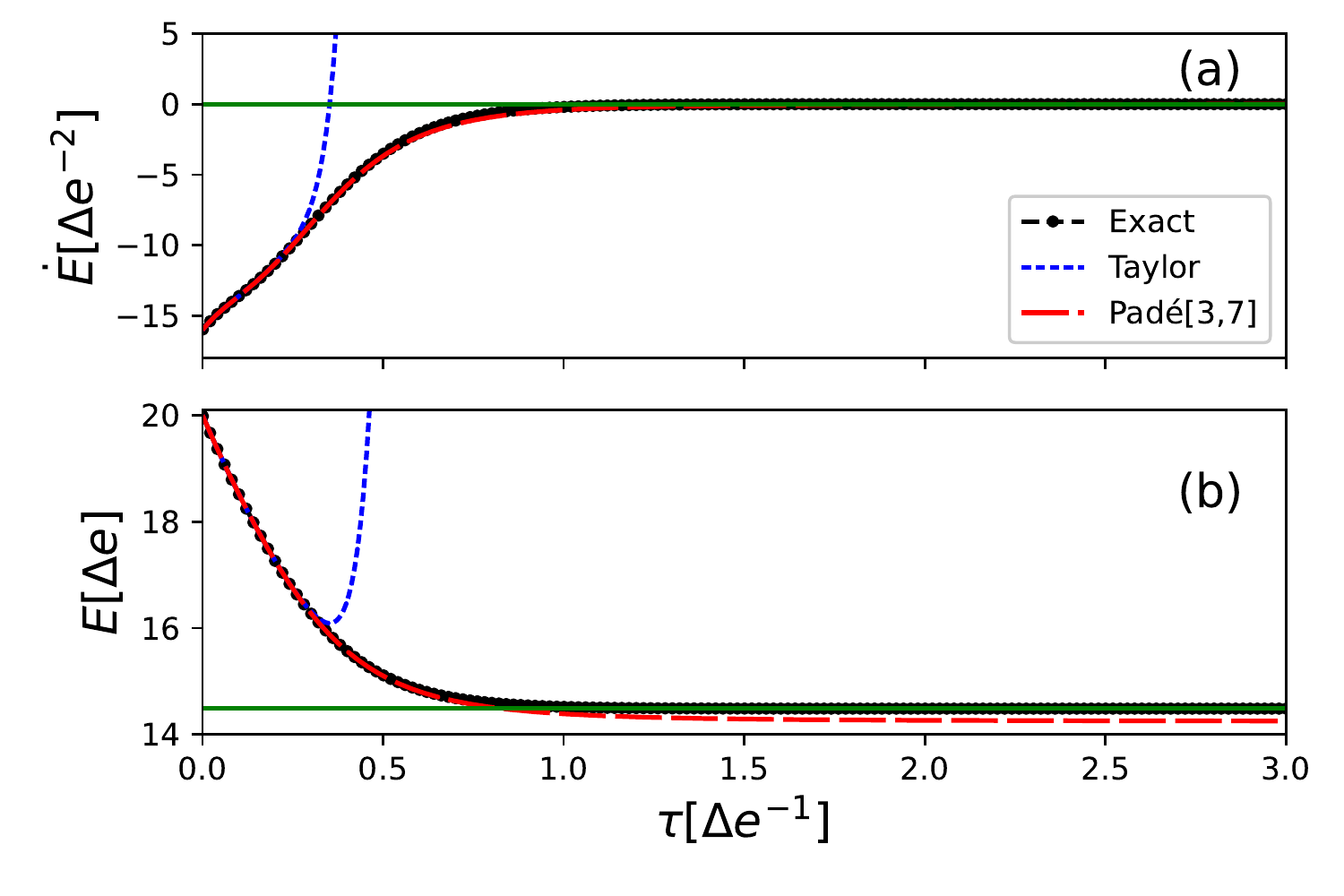}
\par\end{centering}
\centering{}\caption{Illustration of the $t$-expansion method applied to the pairing model.
The simulations are performed for the same conditions as panel (a)
of Fig. \ref{fig:genhubpair}. The derivative of the energy and the
energy are shown as a function of the imaginary time in panels (a)
and (b) respectively. In each panel, the exact (black filled circles),
Taylor expansion of Eq. (\ref{eq:cumexp}) with  $M=10$ (blue short-dashed line)
and the Pad\'e$[3,7]$ (red long-dashed line) are shown. Note that $M=10$ means that
we have used $M+2=12$ cumulants as inputs.    
The exact imaginary-time solution
(black filled circle) was obtained by performing explicitly the imaginary-time
evolution on a classical computer. In panel (b), the green horizontal
line corresponds to the exact ground state energy. 
}
\label{fig:padepairing} 
\end{figure}

As was noted in Ref. \cite{Hor84}, it might be more efficient to
consider the derivative of $E(\tau)$ with respect to $\tau$ than
the energy itself to extrapolate the asymptotic value of the energy.
Here, we follow closely the prescription proposed in Ref. \cite{Hor84}.
The evolution of the energy is given by: 
\begin{eqnarray}
\frac{d}{d\tau}E(\tau) & = & -\left(\langle H^{2}\rangle_{\tau}-\langle H\rangle_{\tau}^{2}\right), \label{eq:dedtau}
\end{eqnarray}
where we introduced the notation $\langle.\rangle_{\tau}$ for the
expectation values taken at time $\tau$ with $|\Psi(\tau)\rangle$.

Assuming that only the lowest $M+2$ cumulants (or
moments) of the Hamiltonian are known, this derivative is approximated as 
\begin{eqnarray}
\frac{d}{d\tau}E(\tau) & \simeq & -\sum_{K=0}^{M}\frac{(-\tau)^{K}}{K!}\kappa_{K+2}.\label{eq:cumexp}
\end{eqnarray}
We then replace this approximate form by a Pad\'e approximation, denoted
by {Pad\'e}$[I,J](\tau)$ where $I$ and $J$ are the orders of the
numerator and denominator respectively. The Pad\'e is adjusted such
that it reproduces the Taylor expansion given above with the constraint
$I+J=M$. The great advantage of using the derivative of the energy
stems from the expression (\ref{eq:dedtau}). Besides the fact that
the derivative tends to zero if the Hamiltonian is bound from below,
we observe that the energy is always decreasing in imaginary-time
evolution. This gives strong constraints on the Pad\'e approximation
that could be used. Due to the fact that the derivative, once integrated
in time, should give a convergent energy, the decrease of the derivative
towards zero should be faster than $1/\tau$. This gives the additional constraint
$J-I\ge2$. Once the Pad\'e approximation that fulfills all these constraints
is obtained, the energy $E(\tau)$ is deduced simply by integrating
the derivative with respect to $\tau$. We have found that, in the
two models considered here, the method is rather accurate to predict
the ground state energy $E_{{\rm GS}}$. 

\begin{figure}[htbp]
\includegraphics[width=1.0\linewidth]{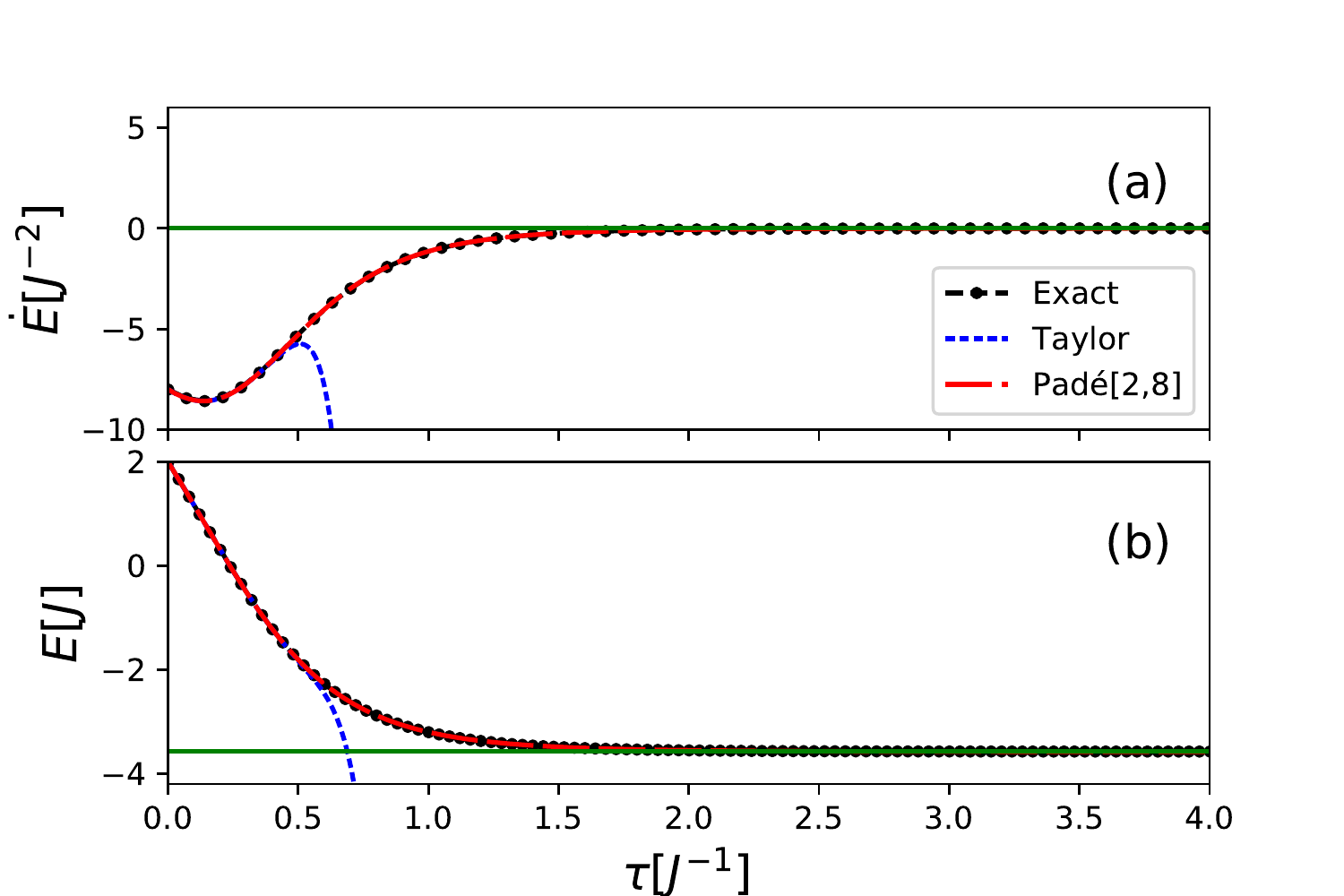}
\caption{Same as figure Fig. \ref{fig:padepairing} for the Fermi-Hubbard model,
with the parameters of Fig. \ref{fig:genhubpair}. Note that, if only
one of the Slater determinants is used instead of the mixing of 6
of them as initial condition, the convergence towards the ground
state energy requires to include higher order moments. }
\label{fig:padehubbard} 
\end{figure}

We illustrate in Figs. \ref{fig:padepairing} and \ref{fig:padehubbard}
the $t$-expansion method applied to the two models. 

We see in both cases that, despite the fact that a finite number of
moments are used and that a truncated Taylor expansion (TTE) can only
describe the short imaginary-time evolution, the method gives results
that are very close to the exact imaginary-time evolution. We clearly
see in the TTE in Figs. \ref{fig:padepairing} and \ref{fig:padehubbard}
that the knowledge of the moments up to a given order only allows
us to reproduce the exact evolution of $\dot{E}(\tau)$ and $E(\tau)$
over a certain interval $[0,\tau_{{\rm max}}]$. 
We show in the illustration here,
that the $t$-expansion method can be used to provide rather accurate
extrapolation of the system's ground state energy.

\begin{figure}[htbp]
\includegraphics[width=1.0\linewidth]{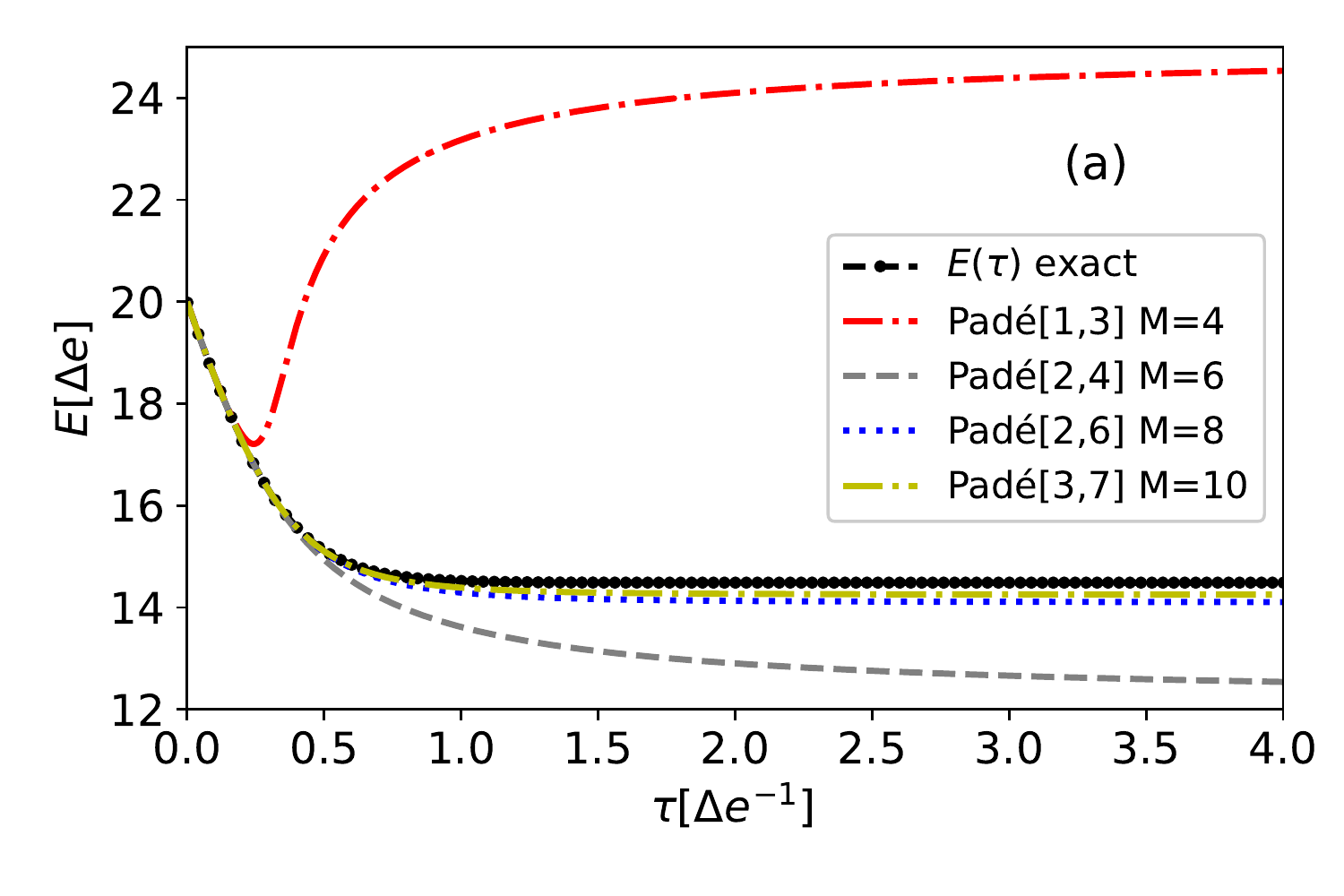}
\includegraphics[width=1.0\linewidth]{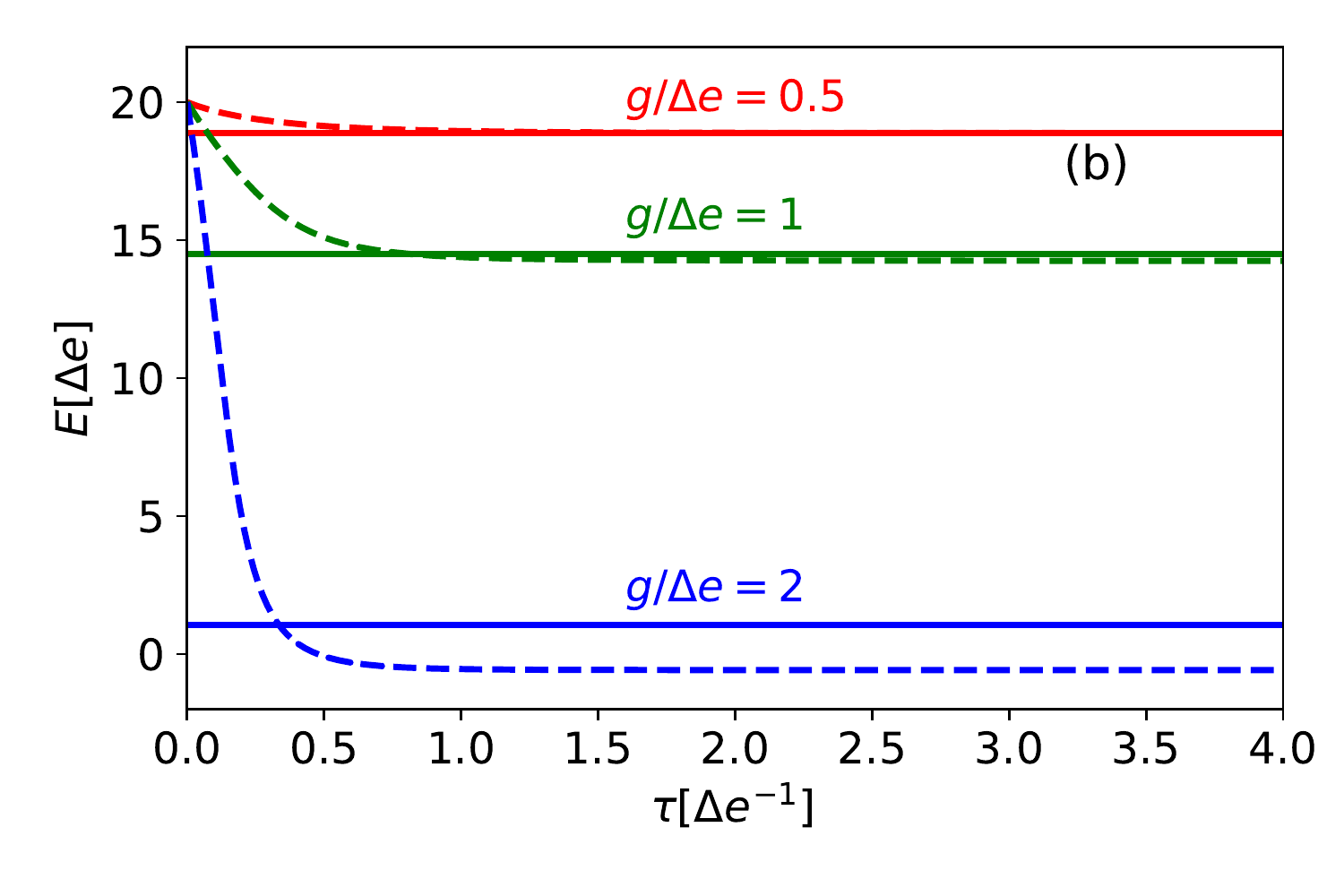}
\\
\caption{Illustration of the convergence properties of the $t$-expansion for
the pairing problem with the same parameters and initial condition
as in Fig. \ref{fig:genhubpair}. In panel (a), results obtained by
changing the orders $(I,J)$ in Pad\'e{[}I,J{]} are presented. Note
that this corresponds to changing the order of truncation $M$ used
for the Pad\'e approximation. For comparison, we also show the result
of the exact imaginary time evolution (black filled circles) that
converges to the exact ground state energy. In Panel (b), 
the results obtained when $g/\Delta e$ is equal to 0.5 (red), 1.0
(green) and 2.0 (blue) are shown. The dashed lines correspond to the Pad\'e{[}3,7{]},
Pad\'e{[}3,7{]} and Pad\'e{[}2,8{]} obtained in all cases with $M=10$
for $g/\Delta e=0.5$, $1.0$ and $2.0$ respectively.}
\label{fig:padepairingsens} 
\end{figure}

The interval of time over which the TTE is valid depends on the order
$M$ of the truncation in Eq. (\ref{eq:cumexp}). It will also be influenced 
by the initial state that is used. We illustrate these two aspects for the pairing case in panels (a)
and (b) of Fig. \ref{fig:padepairingsens}. In panel (a), the dependence
of the convergence of the approach on the number of moments $M+2$
in the inputs is shown (note that the orders $[I,J]$ are changed
accordingly to fulfill the constraints $I+J=M$). 

This panel illustrates the rapid convergence of the method when $M$
increases. Note that the $M=4$ case leads to a very bad asymptotic
value because the only possibility for the Pad\'e in this case (Pad\'e{[}1,3{]})
has a pole leading to a positive unphysical approximation for the
derivative of the energy. This problem disappears when $M$ is increased.
When $M$ is sufficiently high, there is a flexibility in choosing
the order $(I,J)$ even with the constraints given above. We have
empirically observed that higher ratios $I/J$ give better results
than the case $J-I=2$. Another strong guidance, already noted in
Ref. \cite{Hor84}, is given by the fact that the Pad\'e approximation
of $dE(\tau)/d\tau$ should always be negative.

To illustrate the importance of the initial state on the convergence,
we have progressively increased the two-body interaction strength
$g$ while keeping the initial state unchanged. When the strength increases,
this initial state deviates more and more from the exact ground state.
In panel (b) of Fig. \ref{fig:padepairingsens}, we compare the solution
of the $t$-expansion approach with a fixed value $M=10$
with the ground state energy. As expected, the predictive power of the method
degrades with the increase of $g/\Delta e$. It is still rather encouraging to observe
that even for the largest $g$, the result remain reasonably close to
the exact solution. Indeed, above $g/\Delta e=1$,
the pairing problem becomes highly non-perturbative and a good solution
of this problem can only be obtained by using a symmetry breaking
state followed by a symmetry restoration \cite{Duk04,Lac12,Deg16,Rip17}.
We anticipate that the use of initial states obtained using the variation
of projections of a $U(1)$ symmetry broken state, like
the BCS ansatz, will strongly improve the ground state energy prediction
from the $t$-expansion. Work is actually in progress to combine the
two techniques on quantum computers. Although we only explore the Pad\'e technique 
in the present work, we note that the connected moments expansion (CMX) \cite{Cio87} can be used as an alternative method to obtain the ground state 
energy \cite{Cla21,Pen21}.  

\subsection{Excited states and time-dependent evolution}

\label{sec:excandtd}

Starting from an initial state $|\Phi_{0}\rangle$, the real-time
evolution in Hilbert space is given by: 
\begin{eqnarray}
|\Phi(t)\rangle & = & \left(\sum_{K}\frac{(-it)^{K}}{K!}H^{K}\right)|\Phi_{0}\rangle.\label{eq:expprop}
\end{eqnarray}
We recognize in the expansion the Krylov states denoted by $|\Phi_{K}\rangle\equiv H^{K}|\Phi_{0}\rangle$.
In the following, we will consider the Krylov subspace, denoted by
${\cal H}_{M}$, associated to the non-orthogonal basis $\{|\Phi_{0}\rangle,H|\Phi_{0}\rangle,\cdots,H^{M}|\Phi_{0}\rangle\}$.
Note that with the present convention, ${\cal H}_{M}$ contains $(M+1)$
states.

The Krylov basis and Krylov subspace is at the heart of several famous
algorithms to diagonalize sparse matrices \cite{Saa11}. Among the
most popular, we mention the Lanczos and the Arnoldi iterative methods
that are widely used on classical computers. Quantum equivalents to
the Lanczos algorithm have attracted recently special attention \cite{Bes20,Mot20,Bak20,Par19}.
In a sense, the Krylov basis can be seen as an optimal basis to describe
the evolution of a system due to the expansion (\ref{eq:expprop}).
In the absence of truncation of the Krylov basis, we can describe
exactly the evolution for all time. If we now consider the truncated
Hilbert space ${\cal H}_{M}$, we will be able to describe exactly
the evolution up to the order $t^{M}$ of the expansion.

The expectation values of the initial moments of $H$ contain important
information on the Krylov basis. To illustrate the connection between
moments and states, let us restrict the evolution of the system in
a given subspace ${\cal H}_{M}$. Then, we can write the evolution
as: 
\begin{eqnarray}
|\Phi(t)\rangle & = & \sum_{K=0}^{M}c_{K}(t)|\Phi_{K}\rangle,
\end{eqnarray}
with the initial condition $|\Phi(0)\rangle=|\Phi_{0}\rangle$.

The approximate evolution in the subspace ${\cal H}_{M}$ is obtained
by minimizing the time-dependent variational principle: 
\begin{eqnarray}
\delta\int_{0}^{t_{f}}dt\langle\Phi(t)|i\partial_{t}-H|\Phi(t)\rangle dt=0
\end{eqnarray}
with respect to all possible variations of the $c_{K}(t)$ or $c_{K}^{*}(t)$.
From the variational principle, we deduce the set of time-dependent
coupled equations (for all $L$): 
\begin{eqnarray}
i\sum_{K}O_{LK}\frac{dc_{K}(t)}{dt} & = & \sum_{K}H_{LK}c_{K}(t),\label{eq:coupled}
\end{eqnarray}
with the initial condition $C_{K}(0)=\delta_{K0}$. In this equation,
we have defined the matrix elements of the overlap and Hamiltonian
matrix: 
\begin{eqnarray}
\left\{ \begin{array}{l}
{\displaystyle O_{LK}=\langle\Phi_{L}|\Phi_{K}\rangle=\langle H^{K+L}\rangle_{0},}\\
\\
{\displaystyle H_{LK}=\langle\Phi_{L}|H|\Phi_{K}\rangle=\langle H^{K+L+1}\rangle_{0}}
\end{array}\right..\label{eq:ohmoments}
\end{eqnarray}

The equations (\ref{eq:coupled}) correspond to the standard time-dependent
coupled equations (TDCE) that are obtained in a non-orthogonal basis.
We see from the definitions (\ref{eq:ohmoments}) that all the ingredients
needed to solve these equations are linked to the initial moments
of $H$. More precisely, the solution of the TDCE in the subspace
${\cal H}_{M}$ requires the knowledge of the first $L=2M+1$ moments.
We show in the appendix \ref{sec:tdcevalidity} that the use of the
variational principle insures that the approximate solution also matches
the exact evolution up to order $t^{M}$.

The TDCE can be solved by integrating numerically the time-dependent
equations of motion (\ref{eq:coupled}). Alternatively, one can transform
the problem into an eigenvalue problem in the ${\cal H}_{M}$ subspace,
where, for each values of $M$, we generate a set of eigenvalues $E_{\alpha}^{(M)}$
associated to eigenstates denoted by $|\alpha^{(M)}\rangle$. 
Technically, the solution of the problems is equivalent to an eigenvalue problem in 
the non-orthogonal basis formed by the states $\{ |\Phi_{K}\rangle\}$. This problem is rather standard and 
can be solved in two steps: (i) first, the overlap matrix given by the $O_{LK}$ in Eq. (\ref{eq:ohmoments}) is
diagonalized to obtain a new set of ortho-normal state vectors. The hamiltonian is then diagonalized in the new basis.  
We illustrate below an application of this techhnique.

\subsubsection{Excited states from moments}

\begin{figure}[htbp]

\centering{}\includegraphics[width=0.8\linewidth]{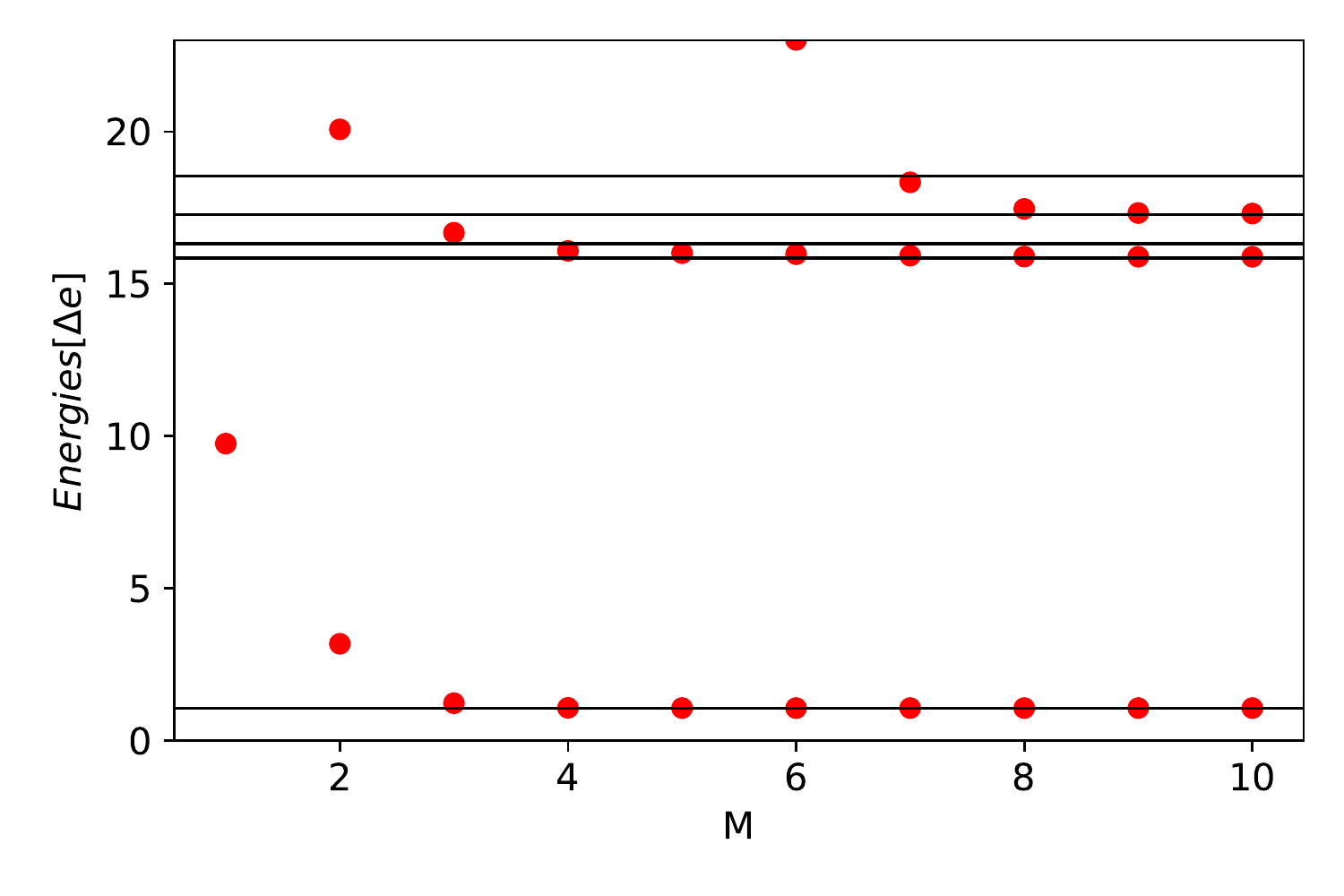}
\caption{Illustration of the eigenvalues (red filled circles) evolution obtained
by the diagonalization of the Hamiltonian in the truncated Krylov basis
for increasing $M$ for the pairing model with the
same conditions as in Fig. \ref{fig:genhubpair} except that the pairing
strength is set to $g/\Delta e=2.0$. Note that the number of associated
moments used as inputs are given by $L=2M+1$. The horizontal black
lines indicate the lowest exact eigenvalues of the pairing Hamiltonian. }
\label{fig:excited} 
\end{figure}

We show in Fig. \ref{fig:excited} the evolution of the $\{E_{\alpha}^{(M)}\}$
values as a function of $M$ for the pairing Hamiltonian case and
for the strong coupling regime $g/\Delta e=2.0$. 

In this figure, we see that the energies obtained by diagonalization
of the Hamiltonian $H_{M}$ with increasing $M$ converge to some
of the exact eigenvalues. The lower is the energy, the faster is the
convergence. For the ground state, we observe that a good accuracy
is already observed for $M=3$ which corresponds to considering the
first 7 moments. In particular, for a number of moments that is lower
than the one used in Fig. \ref{fig:padepairing} for $g/\Delta e=1$,
a much better accuracy is achieved. Note that in general the dimension
of ${\cal H}_{M}$ is rather small compared to the total size of the
Hilbert space ($70$ for the pairing model with 4 particles on
8 levels with zero seniority). We systematically
observed with the two models that the diagonalization method, compared
to the $t$- expansion, not only give access to excited states but
also seems to converge more rapidly to the ground state when the number
of moments increases. Finally, we note that some excited states are
missed due to the fact that their overlaps with the initial state is too low or the Krylov
basis size should be further increased.
By exploring different initial states, one could expect to obtain
the eigenstates that are not reproduced in Fig. \ref{fig:excited}.

\begin{figure}[htbp]

\centering{}\includegraphics[width=1\linewidth]{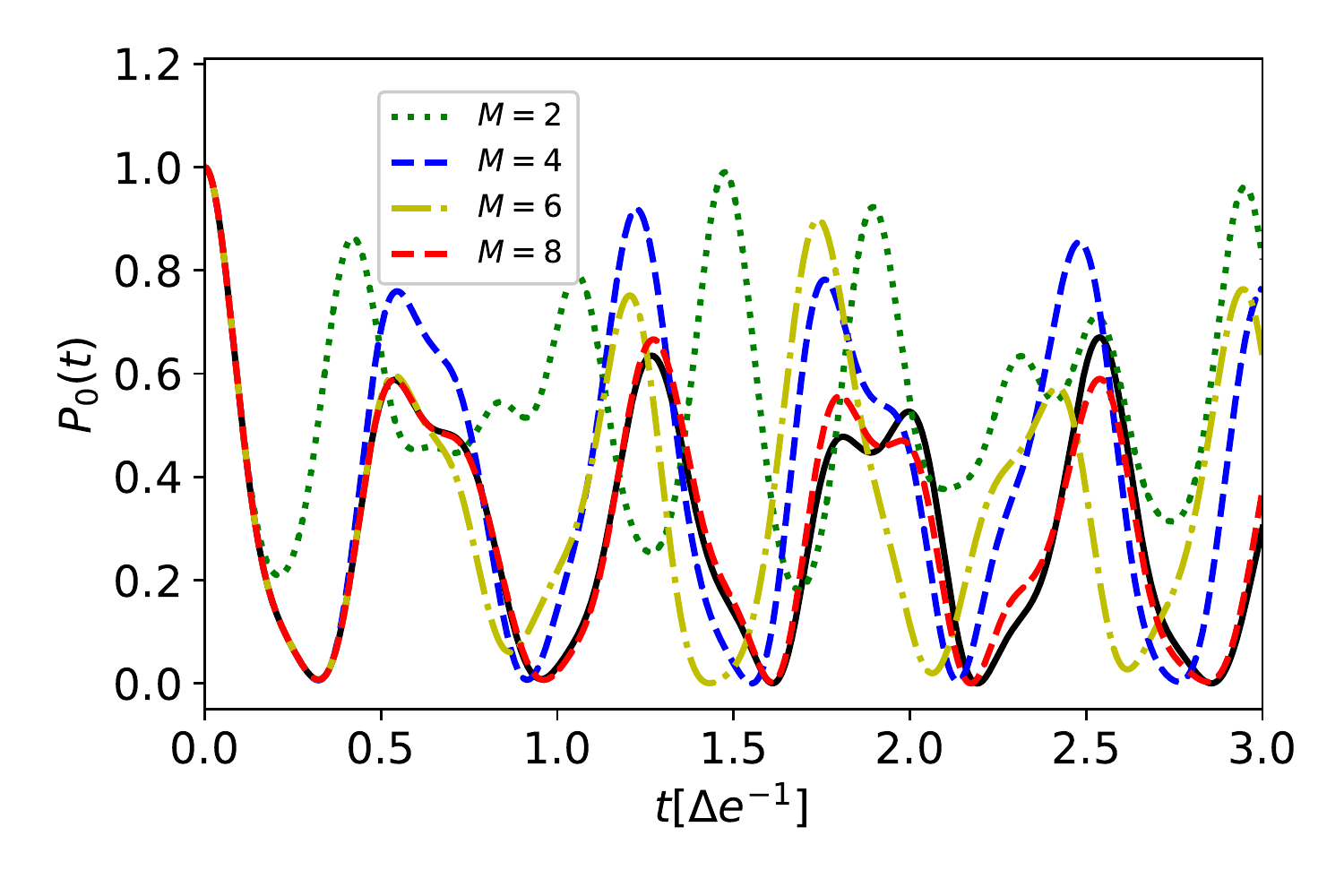}
\caption{Evolution of the quantity $P_{0}^{(M)}(t)=|\langle\Phi_{0}|\Phi^{(M)}(t)\rangle|^{2}$
as a function of time obtained by solving the TDCE equations with
increasing values of $M$ for the pairing model with $g/\Delta e=2$.
The black line shows the exact evolution. }
\label{fig:td} 
\end{figure}

\subsubsection{Long-time evolution from moments}

We now return to one of the main motivations of the present work, i.e.
predict the long-time evolution of a quantum complex system. As a
follow up of the previous section, we now use the $M$ states obtained
by the diagonalization of $H_{M}$. The evolution of the system in
the Hilbert space ${\cal H}_{M}$ is given by: 
\begin{eqnarray}
|\Phi^{(M)}(t)\rangle & = & \sum_{\alpha=0}^{M-1}e^{-iE_{\alpha}^{(M)}t}|\alpha^{(M)}\rangle\langle\alpha^{(M)}|\Phi_{0}\rangle.
\end{eqnarray}
From this, we can compute the evolution of the survival probability
$P_{0}^{(M)}(t)=|\langle\Phi_{0}|\Phi^{(M)}(t)\rangle|^{2}$. Illustrations
of the different evolutions of the survival probability obtained with
different values of $M$ are shown in Fig. \ref{fig:td} and
compared to the exact solution. In all cases, the approximate evolution
matches the exact solution up to a certain time $t_{{\rm max}}(M)$.
This time increases with $M$. This is expected since the method is
designed to give the correct Taylor expansion (\ref{eq:expprop})
of the evolution up to order $t^{M}$. We see also that the evolution
converges towards the exact solution when $M$ increases even if the
number of states included is much lower compared to the size of the
complete Hilbert space. 

It is worth mentioning that if we now make the Fourier transform of the
survival probability to obtain the strength function, already at $M=5$,
one would have a good reproduction of several dominant frequencies.
This is consistent with Fig. \ref{fig:excited} where some of the
exact eigenvalues are already well reproduced at rather low $M$ values.

\subsection{Application on noisy quantum platforms}

As a test, we have tried to compute the generating function on some of the 
real quantum processor units (QPU) available on the IBM quantum cloud. 
We focus here on the specific case of the  {\sl Santiago} QPU. Since the number of 
qubits is limited to 5 in this case, we considered the simple pairing case where 
a single pair of particles can access two different single-particle levels with spacing $\Delta e$. 
Such case can be encoded on 2 qubits, plus an extra ancillary qubits to perform the Hadamard 
or modified Hadamard tests shown in Figs. \ref{fig:hadgen}. Raw results obtained with the 
{\sl Santiago} QPU turn out to be strongly polluted by noise. 

\begin{figure}
\includegraphics[width=1\linewidth]{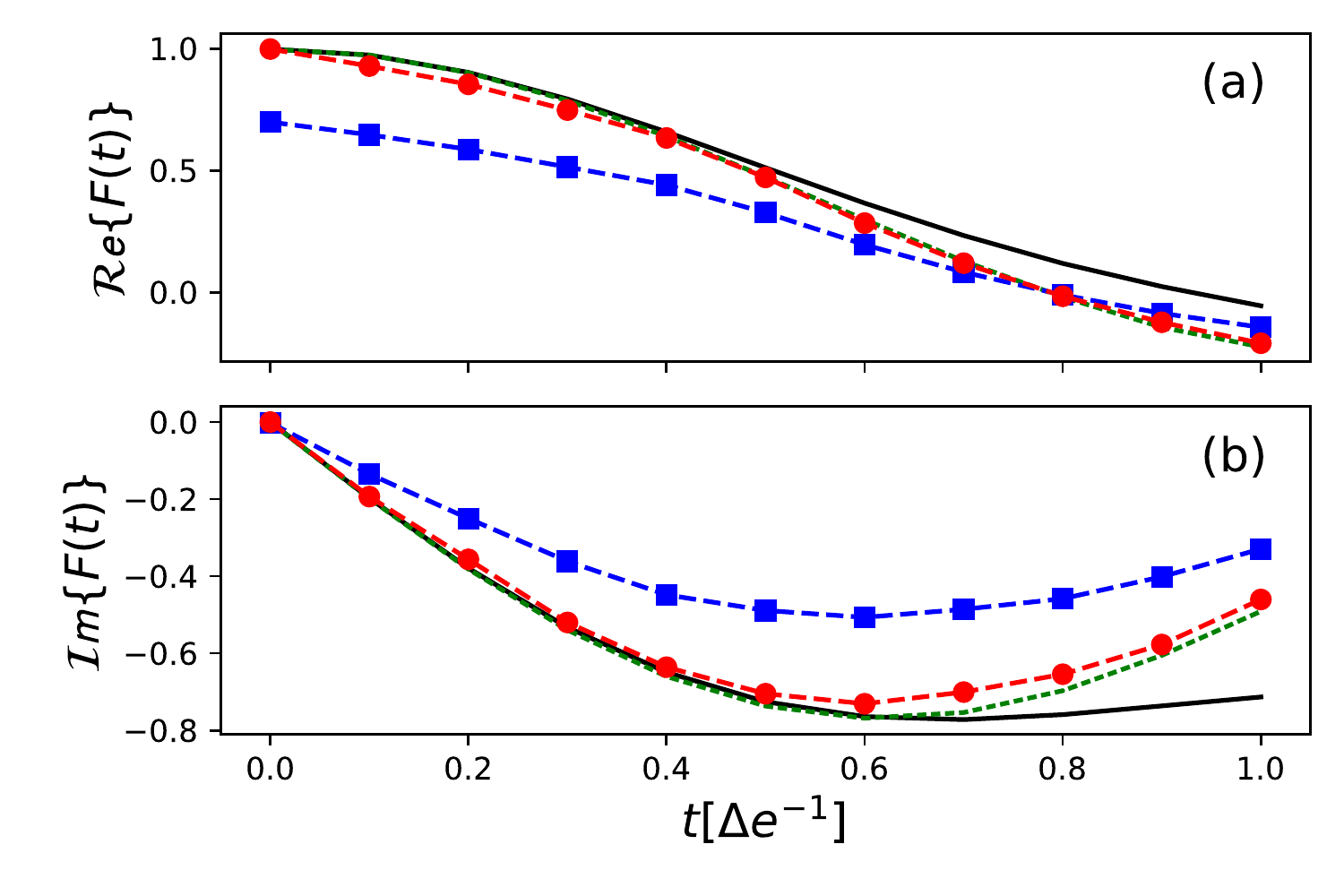}
\caption{Real (a) and imaginary (b) parts of the generating function obtained for the case of 
a single pair on two levels with spacing $\Delta e$ and coupling strength $g/\Delta e = 1$. 
The exact result obtained on a classical computer is shown in black solid line. Results obtained 
with {\sl FakeSantiago} backend without and with error corrections are shown respectively by 
blue squares and red circles. We also show for comparison the results obtained with QASM backend (no noise reference quantum calculation)  
in green long-dashed line. In the quantum simulations, each point is obtained using $10^6$ measurements.}
\label{fig:fakesantiago}
\end{figure}

We therefore have tried to implement some standard noise correction techniques. In order to test 
these error corrections, we have used the {\sl FakeSantiago} QPU that simulates the topology and the noise 
of the real {\sl Santiago} QPU using depolarizing, thermal relaxation and read-out errors. 
An important aspect to notice is that  the implemented circuits on the real and fake {\sl Santiago} QPU were different 
compared to the ones shown in Figs. \ref{fig:hadgen}. This is because these devices have a set of basic gates which can be implemented 
and because each device has its own topology. As {\sl FakeSantiago} emulates the behavior 
of real Santiago they have the same set of basic gates and topology.
The basic gates of {\sl FakeSantiago/Santiago} did not contain all the gates in the circuits of Figs. \ref{fig:hadgen}. Thus the circuits in Figs. \ref{fig:hadgen} had to be replaced
by equivalent circuits that use the set of basic gates of each particular device. Also, the new equivalent circuit has to take into account
the topology of the device in which is being implemented. The complete process of replacing the theoretical circuits with circuits that we can use in the devices is called transpiling.
Qiskit offers several optimizations which can reduce the depth of the transpiled circuits. The results that are shown in Fig. \ref{fig:fakesantiago}
were obtained with a level of optimization of 2 which corresponds to a medium level of optimization. Due to the types of optimizations that this level performs, we can find that different transpilations of the same circuit can generate circuits of different depths. 
In order to address this, we transpiled the circuit for each point 50 times, and implement the one that had the lower depth. We used the 2$^{nd}$ level of optimization because we did not find further improvement when using the 3$^{rd}$ level.

We show in Fig. \ref{fig:fakesantiago}
the evolution of the real and imaginary parts of the generating function obtained with and without the noise. Results without noise 
correspond to the evolution obtained on a classical computer and on the perfect QC emulator (i.e qasm backend) of Qiskit. 

We clearly observed in this figure that both real and imaginary parts deviate quite significantly from the 
exact solution, even at very short time.  These deviations stems from two sources (i) the noise that is added in 
 {\sl FakeSantiago} to simulate the real device and (ii) the discretization of time that was used in the Trotter-Suzuki method. 
 Results obtained in Fig. \ref{fig:fakesantiago} are calculated by simply assuming a single step in the Trotter-Suzuki technique, i.e. 
 for a given time $t$, the time $\Delta t$ of evolution is directly equal to $t$.  This was done in order to minimize the depth of the circuit and thus, the effect of the noise. While for short time $t$, this approximation can be accurate, a single-step 
 in the Trotter approximation will induce deviations from the exact solution when $t$ increases. To illustrate this, we also show in this figure the result obtained 
 with the QASM backend with no noise, same Trotter-Suzuki time-step and same number of measurements. We see that, even in the absence of noise, some deviation 
 with the exact solution occurs when $t$ increases. A simple solution to this problem is to increase the number of steps $n$ in the Trotter-Suzuki methods leading to $t/\Delta t = n$. A drawback is that the depth of the circuit strongly increases when $n$ is increased even by a single unit. This induces a significant increase of the errors on the generating function that could in general not be corrected by the methods 
 discussed below. The results obtained for $n> 1$ with error corrections turns out to be worst compared to the case $n=1$.
 
 As an illustration of the effect of error correction, we show in Fig. \ref{fig:fakesantiago} the results obtained after some specific corrections. 
 To obtain the corrected results, we have  used  several corrections methods, including the read-out corrections of Ref. \cite{Bla04}, supplemented 
 by the  post-selection correction correction of Ref. \cite{Sun20}. Our aim is not here to make a full description of the error mitigation techniques and readers interested in the technical details can refer to the original articles. These two methods correct partially the noise observed in Fig. \ref{fig:fakesantiago}. To further improve the result, we have also adapted the "reference correction" 
 technique proposed also in \cite{Sun20}. In this approach, we use the fact that we already know the values of the generating function 
 at time $t=0$. With this, we can construct a matrix $M$ that connects the noiseless measurements to the real measurements at this time. It is then 
 assumed that the same matrix $M$ applies at all times. Results obtained using the combination of these three error corrections are shown 
 with red circles in Fig. \ref{fig:fakesantiago}. We see that, with these methods, the error made in the {\sl FakeSantiago} device can be rather accurately corrected. 
  
We finally mention that we also tried to apply the same protocol with the real {\sl Santiago} device but the results were more noisy than 
on the fake device and we were not able to obtain reasonable corrected results. This suggest that, in the NISQ period, 
the present approach should probably still be combined with variational technique as explored in Ref. \cite{Cla21,Pen21}.

\section{Conclusion}

We discuss here the possibility to compute the generating function of an operator 
using quantum computers, with a focus on the case where the operator is the Hamiltonian 
itself. The quantum method that we use is based on standard Hadamard tests and is expected
to minimize the quantum resources by using a single ancillary qubit. 
The generating function gives a priori access to the different moments of the 
operators under interest, that are difficult to compute directly on a quantum computer. \\
Provided that the moments could be efficiently computed from the calculated generating function, 
we discuss how this information can be exploited in a post-processing step on a classical
computer. We show that the $t$-expansion method in combination with
Pad\'e approximation can be used to obtain rather accurate estimates
of the ground-state energy. We then illustrate the connection between
the moments and the approximate evolution of the system in a truncated
Krylov space. The latter approach could be used to study ground state
and excited states properties as well as to extrapolate the short-time
evolution performed on a quantum computer to an approximation of
the long time evolution on a classical computer given that we can approximate
with high precision the value of the moments. 
We note finally the recent Ref. \cite{Aul21}, 
where methods based on moments, including the one discussed here, have been discussed. 

One critical aspect to be able to use the generating function as a generator of the moments
is definitely the accuracy achieved in computing the different moments. We actually encountered
significant difficulties in obtaining the $\langle H^K \rangle$ values with high precision when $K$ increases. 
At present, we have not found a better 
solution to this problem than performing the Fourier transform/spectral analysis of the time-dependent generating function.  
Such Fourier transform is rather demanding in terms of quantum resources and, although there is a gain compared to the QPE in 
terms of circuit length, the numerical effort remains quite significant. After all the tests we made, we believed 
that the method based on the calculation of moments from the generating function will be rather hard to apply in the NISQ context 
unless a method, alternative to the Fourier transform and with lower global quantum cost, is found.

\section*{Acknowledgments}

This project has received financial support from the CNRS through
the 80Prime program and is part of the QC2I project. We acknowledge
the use of IBM Q cloud as well as use of the Qiskit software package
\cite{Abr19} for performing the quantum simulations.

\appendix



\section{Validity of the solution in the ${\cal H}_{M}$ space in the TDCE approach}

\label{sec:tdcevalidity}

In the present section, we give a direct proof that the use of the
TDCE equations (\ref{eq:coupled}) in the truncated subspace ${\cal H}_{M}$
insures that the evolution is exact up to order $t^{M}$ in the Taylor
series (\ref{eq:expprop}). In the following, we will denote by $|\Phi^{(M)}(t)\rangle$
the state obtained by solving the TDCE equation and by $|\Phi(t)\rangle$,
the exact solution in the full Hilbert space.

The approximate evolution of the wave-packet associated at a given
order $M$ is given by: 
\begin{eqnarray*}
i\frac{d}{dt}|\Phi^{(M)}(t)\rangle & = & i\sum_{J=0}^{M}\dot{c}_{J}(t)|\Phi_{J}\rangle.
\end{eqnarray*}
Introducing the inverse of the overlap matrix, we can rewrite this
equation as: 
\begin{eqnarray}
i\frac{d}{dt}|\Phi^{(M)}(t)\rangle & = & \sum_{J=0}^{N}|\Phi_{J}\rangle\sum_{KL}O_{JL}^{-1}H_{LK}c_{K}(t)\nonumber \\
 & = & \sum_{JL}|\Phi_{J}\rangle O_{JL}^{-1}\langle\Phi_{L}|H|\Phi^{(M)}(t)\rangle\nonumber \\
 & \equiv & P_{M}H|\Phi^{(M)}(t)\rangle\nonumber \\
 & = & H_{M}|\Phi^{(M)}(t)\rangle.\label{eq:projH}
\end{eqnarray}
In the last equation, we have introduced the projector of the Krylov
subspace ${\cal H}_{M}$ that is given by: 
\begin{eqnarray}
P_{M} & = & \sum_{I,J=0}^{M}|\Phi_{I}\rangle O_{IJ}^{-1}\langle\Phi_{J}|.\label{eq:projM}
\end{eqnarray}
We note in particular that, for all Krylov states with $J\le M$, we
have $P_{M}|\Phi_{I}\rangle=|\Phi_{I}\rangle$. This implies at all
time $P_{M}|\Phi^{(M)}(t)\rangle=|\Phi^{(M)}(t)\rangle$. We used
this last property to obtain the expression (\ref{eq:projH}), where
we have introduced the Hamiltonian projected on ${\cal H}_{M}$, $H_{M}=P_{M}HP_{M}$.

The equation (\ref{eq:projH}) can be formally integrated as: 
\begin{eqnarray}
|\Phi^{(M)}(t)\rangle=e^{-itH_{M}}|\Phi_{0}\rangle.\label{eq:prophm}
\end{eqnarray}
If we now introduce the difference $\Delta_{M}(t)$ between the exact
and approximate evolutions, we have: 
\begin{eqnarray}
\Delta_{M}(t) & \equiv & |\Phi(t)\rangle-|\Phi^{(M)}(t)\rangle\nonumber \\
 & = & \left[e^{-itH}-e^{-itH_{M}}\right]|\Phi_{0}\rangle\nonumber \\
 & = & \sum_{K=0}^{\infty}\frac{(-it)^{K}}{K!}[H^{K}-H_{M}^{K}]|\Phi_{0}\rangle\label{eq:expdif}
\end{eqnarray}
For $K\le M$, because of the properties of the projector, we have:
\begin{eqnarray*}
H_{M}^{K}|\Phi_{0}\rangle & = & P_{M}H^{K}P_{M}|\Phi_{0}\rangle=P_{M}H^{K}|\Phi_{0}\rangle=P_{M}|\Phi_{K}\rangle\\
 & = & |\Phi_{K}\rangle=H^{K}|\Phi_{0}\rangle.
\end{eqnarray*}
Therefore, all terms with $K\le M$ are strictly zero and the first
non-zero term is proportional to $t^{M+1}$.

\end{document}